# Evaluating the Gouy-Stodola Theorem in Classical Mechanic Systems: A Study of Entropy Generation

R. H. Longaresi*

*Departamento de Física, Química e Matemática - DFQM,*
*Universidade Federal de São Carlos, Sorocaba CEP 18052780, Brazil*

S. D. Campos†

*Applied Mathematics Laboratory-CCTS/DFQM,*
*Universidade Federal de São Carlos, Sorocaba CEP 18052780, Brazil*

(Dated: June 20, 2022)

We propose to apply the entropy generation ($\dot{S}_{gen}$) concept to a mechanical system: the well-known simple pendulum. When considering the ideal case, where only conservative forces act on the system, one has $\dot{S}_{gen} = 0$, and the entropy variation is null. However, as shall be seen, the time entropy variation is not null all the time. Considering a non-conservative force proportional to the pendulum velocity, the amplitude of oscillations decreases to zero as $t$ grows. In this case, $\dot{S}_{gen} > 0$ indicates that it is related to energy dissipation, as stated by the Gouy-Stodola theorem. Hence, as shall be seen, the greater the strength of the non-conservative force, the greater are both the energy dissipation and the time rate of entropy variation.

## I. INTRODUCTION

According to Albert Einstein, thermodynamics is the only theory (of physics) with a universal character[1]. Considering this, we can say that the First Law of Thermodynamics (FLT), elaborated from the evolution of the thermodynamics concepts in the 19th century, is one of the cornerstones of Physics.

In parallel with energy conservation, Rudolf Clausius developed and defined the concept of entropy, yielding the Second Law of Thermodynamics (SLT)[2]. This concept deals with the heat flow in a system (reversible or irreversible), showing that small variations that cannot be quantified accumulate in the system leading to a thermodynamic state unable to do work. In other words, Clausius teaches us the importance of knowing if the energy is conserved in a system and how the quality of energy degrades itself (or not) after each cycle of the system.

* longaresi@ufscar.br
† sergiodc@ufscar.br

The First and Second Laws of Thermodynamics are fundamental for understanding possible forms that energy may take and how it can be used in a physical system. Although both laws usually are seen in university and secondary education textbooks, their use has been restricted to thermal systems, in which heat can be stored and transferred. We know, however, that there was an intense study at the beginning of Thermodynamics on the equivalence between mechanical energy and heat, as demonstrated by the Joule experiments[3]. So, it is fair to ask why there are so few studies about mechanical systems within thermodynamics in physics courses. Because of this scarcity, even simple questions remain unanswered in the classroom, for example: is it possible to obtain the equations that describe the motion of a simple pendulum from the laws of thermodynamics?

Despite the existing gap in studies of mechanical systems, even in thermal systems, entropy production (or entropy generation), a physical concept that has been used even in Cosmology, is not seen in undergraduate physics courses, and only in the last 30 years, this idea has been satisfactorily developed in engineering[4].

It is not surprising that a theorem that appeared in the late 19th-century, which deals exactly with entropy generation in mechanical systems, is almost unknown in physics. This theorem, obtained independently by G. Gouy[5] and A. Stodola[6], deals with entropy generation and relates it to the system useful energy, called exergy. The term exergy is usually used in engineering, referring to the increasing in time entropy rate due to the fact that part of energy in the system cannot be used by the limitation imposed by the SLT.

We intend to present the physical and mathematical aspects for obtaining the main result of the Gouy-Stodola theorem. As an example, we use the simple pendulum of Classical Mechanics (CM) considering a damping force, essential for the existence of entropy in this type of mechanical system. We also show how to obtain from $T\mathrm{d}S$ equation the equations describing the pendulum motion. Thus, from the fundamental equations of thermodynamics, we obtain the equations of motion for the pendulum, showing that the physics behind the phenomenon may have different representations, leading to the same physical description of the system.

This work is organized as follows. In section I, we define some basic quantities within Thermodynamics, taking special attention to the entropy definition. In section II, we present the Gouy-Stodola theorem. Section III deals with the definition of a simple pendulum subjected to a damping force, fundamental for entropy production. Section IV brings the analysis of the simple pendulum thermodynamics. Finally, section V deals with entropy generation in the simple pendulum, according to the Gouy-Stodola theorem. Our final comments and conclusions are left to section VI.

## II. BASIC CONCEPTS

Thermodynamics is a macroscopic theory dealing with experimental results on a large scale. From this theory, FLT defines the energy conservation for any system, whether the system is open or closed, isolated or not. Although energy conservation is a fundamental principle of nature nowadays, it has not always been that way. Known since the Newton epoch, mechanical energy conservation in a non-dissipative system was not seen as essential until the beginning of the 19th century, when experiments on mechanical systems, heat, and heat flux have been performed. Among several results, we highlight the Joule experiment, which elegantly showed the equivalence between mechanical energy and heat[3,7].

Considering a thermodynamic system in equilibrium between any two physical states, FLT establishes that the change in internal energy is given by the difference between the heat and the work done by/on the system. Mathematically, one writes

$$Q = W + \Delta U, \tag{1}$$

where $Q$ represents the amount of heat absorved/liberated in a specific process, $W$ is the work done by/on the system and $\Delta U$ is the change in the internal energy of the system between the two states considered. Here, the amount of heat is a function of the work and the internal energy of the system, $Q = Q(W, \Delta U)$ and, implicitly, it depends on the temperature of the system, $T$, which directly acts in the internal energy $\Delta U = \Delta U(T)$. In the general case, the state function $U$ may depend on other variables of interest besides $T$. We can write, for example $U = U(T, V, N_k, \mu_k)$, being $V$ volume of the system, $N_k$ the number of particles of the species $k$, with chemical potential $\mu_k$. There are also several cases of interest in the literature where the internal energy of the system is replaced by potential energy, with or without explicit dependence on temperature (see Ref.[8]).

As FLT is intrinsically related to the energy conservation law, it also can be applied to non-thermodynamic systems, where the temperature is not a parameter, for example, in systems with quantum or topological phase transitions that occur at $T = 0$ K. Furthermore, in these systems, we should compute the total energy, which may include energy from conservative forces (different definitions of potential energy), the kinetic term, and other possible forms (e.g., the chemical potential). Therefore the term $\Delta U$ should be replaced by $\Delta E$, which corresponds to the change of the total energy $E$ of the system, and the work should correspond to the work of non-conservative forces. Thus for systems in which only conservative forces act, the heat flux corresponds to the total energy of the system. Representing the work of a non-conservative force by $W^*$, we can rewrite the equation for the heat associated with a non-thermodynamic system as[1]

$$Q_{nt} = W^* + \Delta E, \tag{2}$$

noting now that $Q_{nt} = Q(W, \Delta E)$ with no implicit dependency on $T$.

A process is defined as adiabatic when there is no heat transfer between two systems or the system and the environment. For systems in which non-adiabatic processes occur, it is possible by using FLT to determine the maximum work that the system can do (or be done over it). Considering two physical states 1 and 2, the initial and final states of the system, respectively, the heat in the non-adiabatic case can be written as

$$Q = W - W_{ad}, \tag{3}$$

where $W$ is the total work done by the system (or over it) and $W_{ad}$ is the work exclusively due to the adiabatic processes connecting the states 1 and 2. Relation (3) alternatively defines the thermodynamic heat, $\Delta U = -W_{ad}$.

Although equation (2) does not restrict the values of the energy for the initial and final state, the physical system imposes limitations on the amount of heat that can be absorbed or released. That constraint is given by the Second Law of Thermodynamics (SLT), which defines entropy. Historically, Clausius coined this word to refer to what is, perhaps, the inevitable end of energy in irreversible processes: it cannot be transformed into useful work. Notice that Clausius initially called this *transformational property* content in a paper published originally in 1850, and reprinted in 1867[2]. Only in 1865, did he creates the word entropy, which means transformation in Greek[9]. Intentionally, by the way, the word entropy was coined to have similarity with energy, since he believed that the physical meaning of both was so close that it could help in its understanding (see page 357 of Ref.[9]).

---

[1] Notice that for the adiabatic process, the change in the total energy decreases due to the work done by a non-conservative force.

Several texts deal with different definitions of entropy and degrees of depth[7]. Here, we are interested in the Gouy-Stodola theorem[4–6] which concerns not the definition of entropy, but the calculation of its production in an irreversible physical system. In this way, we focus our attention on studying the entropy variation in two conditions: considering the whole system, $\Delta S_{sis}$, and the reservoir contained in this system, $\Delta S_{res}$. Thus, for a system in thermal contact with a single reservoir at a constant temperature,[2] and considering the Kelvin definition for SLT[7], we can write the variation of entropy according to the inequality

$$\Delta S_{sis} + \Delta S_{res} \geq 0, \qquad (4)$$

where the equality is valid only for reversible processes and the inequality sign for irreversible processes. For a process where the reservoir releases a heat flux[3], for which the lost heat can be associated to dissipation, finite temperature difference, or transformation ($Q < 0$), then there is less adiabatic work available than the total work $W$, as can be seen from equation (3). Consequently,

$$\Delta S_{res} = -\frac{Q}{T}, \qquad (5)$$

and, therefore, the entropy of the system is given by

$$\Delta S_{sis} \geq \frac{Q}{T}. \qquad (6)$$

Inequation (6) imposes a limitation on the amount of work that can be performed by/on the system, which is given by equation (3), since the maximum heat flux that can be absorbed/released by the system from/to the reservoir is limited to a minimum change in entropy of the system, assuring the inequality in (4) to irreversible processes. Hence, the change in entropy of the system entropy between any two states should always increase and correspond to the minimum value of $Q/T$ in reversible processes.

---

[2] A thermal reservoir by definition is a system that absorbs/releases an infinite amount of heat without changing its temperature.

[3] The heat flux can be defined as the amount of heat that passes through a given surface in a given time interval.

Here it is necessary to clarify the difference between the First and Second Laws of Thermodynamics. The FLT refers to the definition of energy conservation, as expressed by equation (2): energy can be transformed (or represented) in different ways, and the sum of all these ways remains constant in the system. As it concerns our discussion, FLT is inviolable since we can, in principle, account for every kind of energy in the system, in such a way the energy conservation law is satisfied. The SLT, on the other hand, deals with how energy conversions in the system contribute to the work. Conversion of available energy is carried out in the form of work done by/on the system. However, as the work is defined from a path connecting two physical states, the SLT constrains the processes that the system can achieve considering the work performed. Thus, SLT imposes limits on the use of energy available to the system.

From equations (2) and (6), we can combine the First and Second Laws of Thermodynamics to yield the well-known inequality

$$T\Delta S \geq W^* + \Delta E, \qquad (7)$$

here $\Delta S$ is the change in the entropy of the system, and the inequality sign holds for irreversible processes, which are fundamentals to entropy production. We reinforce that the temperature $T$ is constant in the system under study and represents the temperature of the thermal reservoir.

The irreversibility concept in a thermodynamic system is closely related to the finite difference in the temperature between two (or more) parts of a system. Thus, the heat flow between these bodies has a well-defined direction, such that the body with a higher temperature "releases" heat to the one with a lower temperature. According to SLT, the heat flow in the opposite direction would be impossible since we are not considering work done on the system. Thus, irreversibility applies to thermodynamic systems when we consider the direction of heat flow. For systems in CM, the definition of irreversibility resides in the action of non-conservative forces acting on the system. Therefore, due to these forces, the system is unable to return to the initial state because of energy dissipation. Thus, we can associate the increase in entropy in CM systems with the action of non-conservative forces, which decreases the useful energy amount that can be used to perform work.

## III. ENTROPY PRODUCTION

The fundamental result to compute the entropy generation is the Clausius inequality, which is always present in a thermodynamic cycle[4], expressed by

$$\oint_C \frac{\delta Q}{T} \leq 0 \tag{8}$$

where $T$ is the temperature of the thermal bath and $\delta Q$ is an infinitesimal amount of heat[5] along the C closed path. Note that the relation expressed by (8) establishes a continuous sum of all thermal reservoirs whose heat $\delta Q_i$ occurs at temperature $T_i$. The equality in the above condition is valid for reversible systems, while the inequality applies to irreversible ones. The Clausius inequality can be viewed as a restatement of the Kelvin-Planck definition, which establishes that for any system operating in a thermodynamic cycle, it is impossible to convert all the heat received from the thermal reservoir into work[7].

The inequality (8) allows taking into account the presence of internal irreversibility such as heat transfer due to finite temperature differences, spontaneous chemical reactions, non-conservative forces, inelastic deformations, magnetization, or polarization hysteresis, electric current flow through resistors, etc.[7,10,11]. A more convenient way to study inequality (8) is to treat it as an equation and analyse the behavior of its possible solutions. But what is the physical meaning of this quantity to be created? Keep in mind that inequality (8) concerns the possibility of irreversibility within a thermodynamic system. To facilitate the visualization of this condition, we can rewrite the inequality (8) as

$$\left(\int_1^2 \frac{\delta Q}{T}\right)_{irrev} + \left(\int_2^1 \frac{\delta Q}{T}\right)_{rev} \leq 0, \tag{9}$$

where $C$ is again a closed path, but now it was divided into an irreversible process from state 1 to 2, followed by a reversible process from state 2 to 1 [6]. To transform the inequality (8) into equality, we note that we can always write the Clausius inequality as

$$\oint_C \frac{\delta Q}{T} = -S_{gen}, \tag{10}$$

---

[4] A thermodynamic cycle is the one in which, for a finite number of thermodynamic processes, the system returns to its original state and, hence, the thermodynamic variables of state have null variation.

[5] In this text, we will use the $\delta$ notation to represent an infinitesimal variation of an inexact differential.

[6] For the sake of simplicity, we assume only one cycle constituted of two processes. Of course, one can suppose a finite number of each one, rearranging the terms to separate reversible from irreversible ones.

where $S_{gen}$ is a function defined as the generated entropy along the process that passes through successive states contained in the path $C$. For a process free of any irreversibility, $S_{gen} = 0$, but when there is irreversibility, $S_{gen} > 0$. As defined above, entropy is an extensive quantity: for a system composed of a finite sum of disjoint parts, the total entropy is equal to the sum of the entropy of parts. Of course, entropy can be defined as a nonextensive quantity, as the Tsallis entropy[12], which is not addressed here. From the statistical point of view, entropy is related to the number of possible configurations that the physical states of the system can assume. Then, notice that in a reversible system, the null entropy variation statistically indicates that there is no increase in the possible configurations of the system. On the other hand, for an irreversible process, there is an increase in the possible configurations[7]. However, we do not extend the point of view of the statistical mechanics to the problem, since we are only interested in the thermodynamic approach.

Using equation (9), we can rewrite inequality (10) as

$$\left( \int_1^2 \frac{\delta Q}{T} \right)_{irrev} + \left( \int_2^1 \frac{\delta Q}{T} \right)_{rev} = -S_{gen}. \tag{11}$$

The second term on the left-hand side of the equality (11) is, by definition, the change in entropy between states 1 and 2 along a reversible process, that is,

$$S_1 - S_2 = \int_2^1 \frac{\delta Q}{T}, \tag{12}$$

and, as already mentioned, for the reversible path there is no entropy generation ($S_{gen} = 0$)[8]. Therefore, the entropy generated and accounted for in equation (11) must correspond exclusively to the internal irreversibility present along the path of $1 \to 2$. Thus, rearranging the terms of equation (11), we can interpret the entropy change, or the entropy balance of the system, in the form

$$S_2 - S_1 = \int_1^2 \frac{\delta Q}{T} + S_{gen}, \tag{13}$$

where the left-hand side represents the change in the entropy of the system, which is calculated at the initial and final states. Therefore, it is independent of the process in which states 1 and 2 can be connected. The right-hand side, however, is explicitly dependent on the process and cannot be determined only by knowing the initial and final states. The first term on the right-hand side represents entropy transferred from (or into) the system and is related to the heat transfer from (or into) the system along the irreversible process that connects the initial and final state. Therefore, the direction of entropy transfer due to irreversibility follows the same direction as heat transfer: $Q > 0$ implies $\Delta S_{sis} > 0$, then the system absorbs heat; $Q < 0$ implies $\Delta S_{sis} < 0$, then system releases heat.

---

[7] Entropy can decreases in a system only if work is performed on it. Remember how the refrigerator works.
[8] Of course, we can boldly write $S_1 - S_2 = 0$ since the process is reversible. But, the purpose of equation (12) is to study in what conditions this variation is null, as shall be seen by applying equation (13).

Entropy variation in a system is calculated for both entropy transferred and entropy generated, as shown by the second term on the right-hand side of the equation (13). The term $S_{gen}$ is always positive when there is some internal irreversibility in the process and is null on the contrary case. Therefore, we can interpret SLT from $S_{gen}$: the terms on the right-hand side are null for reversible processes and, therefore, the entropy of the system does not change; for irreversible processes, the entropy of the final state will always be greater than the initial one due to the entropy generation.

Depending on the system features, it is useful to express the entropy balance in terms of a time derivative. Thus, we write from equation (13)

$$\dot{S} = \sum_j \frac{\dot{Q}_j}{T_j} + \dot{S}_{gen}, \tag{14}$$

where $\dot{S}$ represents the time derivative of entropy variation, $\dot{Q}_j/T_j$ is the time derivative of entropy transfer at temperature $T_j$ and, $\dot{S}_{gen}$ is the entropy production rate. For this representation, it is evident that in an isolated system ($\dot{Q}_j = 0$) entropy remains constant for reversible processes, as well as increases as entropy production rises due to internal irreversibility present in an irreversible process.

Although entropy variation should be different when considering reversible and irreversible processes, energy conservation remains valid for both cases and can be computed according to equation (2). For a system containing internal irreversibility, energy dissipated throughout the process between initial and final states should be related to entropy generation in the system for that process. To see that, consider a closed system[9] in thermal contact with a single thermal reservoir at temperature $T$, and that any two equilibrium states have an infinitesimal difference in their extensive properties. The work $\delta W$ done by the system

---

[9] This result is also valid for open systems with mass variation, where the final result is still given by the equation (21).

between two infinitesimal states can be expressed by FLT as

$$\delta W_{rev} = -dE + \delta Q, \tag{15}$$

$$\delta W_{irrev} = -dE + \delta Q', \tag{16}$$

where $\delta Q$ and $\delta Q'$ represent an infinitesimal heat variation for reversible (with $\delta W_{rev}$) and irreversible processes (with $\delta W_{irrev}$), respectively. From equation (13), we can rewrite the equations for the work done by/on the system,

$$\delta W_{rev} = -dE + TdS, \tag{17}$$

$$\delta W_{irrev} = -dE + (TdS - TdS_{gen}), \tag{18}$$

where $T$ represents the temperature of the thermal reservoir at which the heat flows into (or outside) the system. We can calculate the time derivative of the work done by the system in a similar way to the entropy balance. Then,

$$\delta \dot{W}_{rev} = -d\dot{E} + Td\dot{S}, \tag{19}$$

$$\delta \dot{W}_{irrev} = -d\dot{E} + (Td\dot{S} - Td\dot{S}_{gen}). \tag{20}$$

Subtracting equation (20) from the result (19), we obtain the desired equation, written below

$$Td\dot{S}_{gen} = \delta \dot{W}_{rev} - \delta \dot{W}_{irrev}. \tag{21}$$

The above result is known as the Gouy-Stodola theorem[5,6], having been independently obtained by the physicists Georges Gouy[5] and Aurel Stodola[6] in the late 19th century. This theorem presents a practical way of obtaining entropy production in a physical system, qualitative and quantitative. The presence of irreversibility leads to a temporal decrease in energy and therefore $\delta \dot{W}_{irrev} < \delta \dot{W}_{rev}$, agreeing with the fact that $\dot{S}_{gen}$ is always positive. The decrease in energy dissipated by the system is responsible for entropy production. Therefore, through the Gouy-Stodola theorem, it is possible to calculate the entropy production rate by the difference between the work in the reversible and irreversible processes in which the systems pass through. We will illustrate this concept using the simple pendulum, the mechanical system mostly used in physics teaching.

## IV. THE SIMPLE PENDULUM

The simple pendulum is a physical system commonly used to study simple harmonic motion. It consists of a body of mass $m$ suspended by a cord that has an inextensible length $l$ and, often, negligible mass, which oscillates around the equilibrium point. For the sake of simplicity, we do not deal with the physical pendulum, which replaces the string with a rigid rod.

At this moment, we describe the simple pendulum as free of irreversibility (non-conservative forces). Then, the resultant force responsible for the pendulum motion is the conservative force due to the gravitational field, the weight force. This pendulum oscillates according to the following equation of motion[10] (adopting $\theta = \theta(t)$)

$$\ddot{\theta} + \omega^2 \theta = 0, \tag{22}$$

where $g = |\vec{g}|$ and $\omega^2 = g/l$ is its frequency of oscillation. In this system, due to energy conservation, the pendulum motion ceases only when a damping force is applied to the system. Thus, the possibility of inserting irreversibility in the system can be done through non-conservative forces (drag or damping force), which brings us closer to a realistic description of the behavior of the simple pendulum. There are several possibilities for the functional form of the drag force, which can account for different features of the system. Here, we will analyse the simplest case, but one that is easily generalized to different types of force. In this way, we consider the drag force acting on the system given by

$$\vec{F}_{n.cons} = \pm b[\vec{v}(t)]^\nu, \tag{23}$$

where $b$ is the damping or drag factor and it may depend on the physical characteristics of the medium, such as density and temperature, for example. Notice that $b$ has a dimensional dependence with the exponent $\nu$ in a such way that the force $\vec{F}_{irrev}$ is expressed by kgm/s$^2$. The index $\nu$ assumes integer values, with $\nu$ even implying the use of the positive sign and $\nu$ odd, the negative sign. For the sake of simplicity, we will use $\nu = 1$ and, therefore, $b$ has unit of kg/s in the International System. Figure 2(a) shows this system.

It should be stressed that the mechanical energy of the system is dissipated due to damping force, resulting in the motion ceasing. Using the simple relation $-bv(t) = -bl\dot{\theta}(t)$, with $v(t) = |\vec{v}(t)|$, we can introduce the non-conservative force (23) in the equation (22), which gives us the well-known equation of motion

$$\ddot{\theta}^2 + \gamma\dot{\theta} + \omega^2\theta = 0, \quad (24)$$

where $\gamma = b/m$ is the damping parameter that characterizes the system as underdamped, overdamped, and critically damped. This parameter is not present in the irreversibility-free case shown in the equation (22) and is responsible for the decreasing of the pendulum oscillation amplitude (Figure 1). Thus, the greater (smaller) $b$, the greater (smaller) $\gamma$, leading to a smaller (greater) number of oscillations for the same time interval.

---
[10] In this text we consider only small oscillations, where $\sin\theta \approx \theta$.

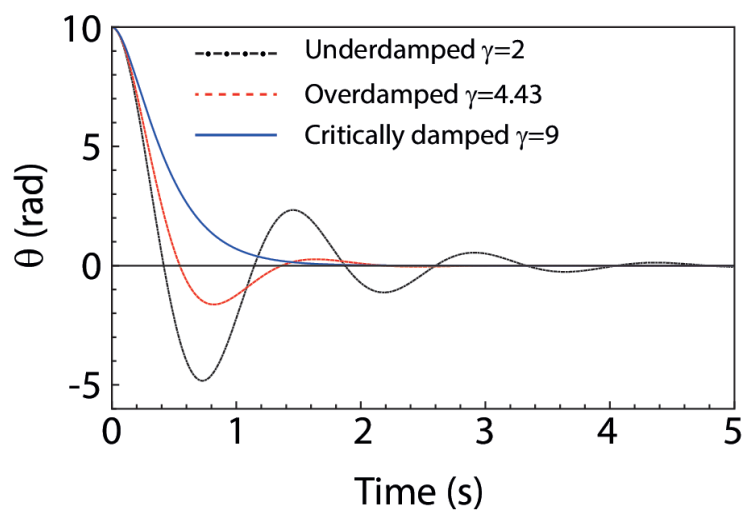

FIG. 1: The behavior of $\theta(t)$ considering the non-conservative force given by equation (23). The oscillatory motion with damping can be characterized as underdamped, overdamped, and critically damped. For calculation purposes, $g = 9.81$ m/s$^2$, $m = 1$ kg, $l = 0.5$ m and $\omega = 4.43$ rad/s.

The irreversibility introduced by the damping force, given by equation (23), leads to a decrease in the total energy due to the work done by the non-conservative force $\vec{F}_{nc}$. This work can be written as

$$W_{irrev} = W_{irrev}(t) = \int_{\theta_0}^{\theta} F_{nc}(\theta')l d\theta' = -bl^2\dot{\theta}(\theta - \theta_0) \quad (25)$$

where $\theta_0 = \theta(t=0)$ is the initial angle. From the above result, one can see that the dissipated energy is responsible for the increase in entropy production rate. It should be noted that we are assuming that the drag force introduced in the system does not change its initial temperature $T$. Therefore, the energy dissipation does not lead to an increase in $T$.

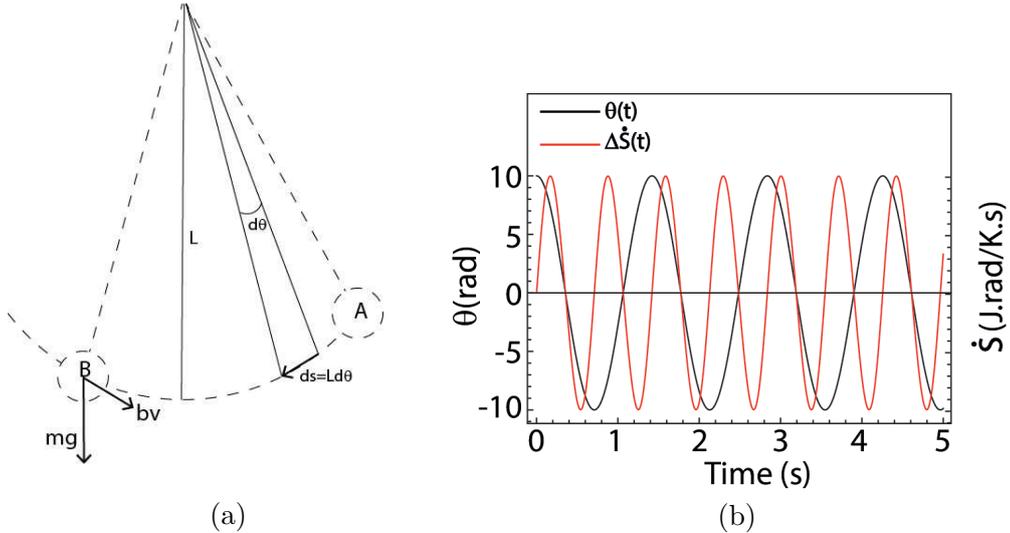

FIG. 2: Panel (a) presents the schematic representation of the simple pendulum. Panel (b) shows the angular position and entropy generation depending on $t$. For calculation purposes, $g = 9.81$ m/s$^2$, kg, $l = 0.5$ m, and $T = 300$ K. Notice that $\theta(t)$ is normalized by the value at the maximum amplitude of $\dot{S}$.

## V. THE SIMPLE PENDULUM THERMODYNAMICS

For the sake of simplicity, this section deals with conservative and non-conservative systems. Furthermore, the physical system studied is composed of the simple pendulum and its surroundings, both at the same constant temperature $T$. Thus, we describe the simple pendulum as a system in thermal contact with a thermal reservoir (neighborhood), both maintained at a constant temperature $T$[11].

### A. Conservative Force

As aforementioned, the pendulum analysed has an initial state A and a final state B, which varies for the non-conservative case, as shown in Figure 2(a). From equation (7) and considering only conservative forces, one writes

$$T(S_B - S_A) = \frac{1}{2}ml^2(\dot{\theta}_B^2 - \dot{\theta}_A^2) + mgl(\cos\theta_A - \cos\theta_B), \qquad (26)$$

that represents a detailed version of $T\Delta S = \Delta K + \Delta U_{grav}$.

---

[11] We consider the thermal bath as the air surrounding the pendulum and maintained at a constant temperature of $T = 300$ K.

The right-hand side of the equation (26) corresponds to a constant value since the system entropy cannot increase or decrease, always reaching the same value in A and B. Thus, one has $S_{gen} = 0$, and for any period of oscillation occurs the transformation of kinetic energy into potential (upward pendulum motion) and potential energy into kinetic (downward pendulum motion). Thus, the reversible system represented here by the simple pendulum is prohibited from reaching any turning points other than A and B. For any two states differing infinitesimally from an angular position $d\theta$ and velocity $d\dot\theta$, we can rewrite the equation (7) in the differential form as $TdS = dE + \delta W$, resulting in

$$TdS = ml^2\dot\theta d\dot\theta + mgl\sin\theta d\theta \tag{27}$$

where, again, one has the condition $dS = 0$, resulting in the right-hand side of above equation can be written as

$$l\dot\theta d\dot\theta = -g\sin\theta d\theta, \tag{28}$$

and assuming the small angle approximation, one has

$$\dot\theta d\dot\theta = -\omega^2\theta d\theta, \tag{29}$$

which is equivalent to the equation (22). To see that, one can integrate both sides

$$\dot\theta[\ddot\theta + \omega^2\theta] = 0, \tag{30}$$

which follows two possible cases. The first one is $\dot\theta = 0$, which means that the change in angular position is constant in time, i.e., the angular velocity is constant, which is not expected. The second case, results in a differential equation (22), which gives us the position of the pendulum according to $t$. Thus, the thermodynamic analysis of the pendulum results in equation (29), which can be obtained by CM, as shown by equation (22).

From the resolution of equation (22) or (29), it is clear that for the maximum amplitude, the velocity achieves its minimum, and for the maximum velocity, $\theta$ is null. From the mathematical point of view, equation (29) allows us to observe that every point whose derivative is zero corresponds to the maximum velocity points, as represented by Figure 4(a). It is important to note that, although we have $\Delta S = 0$ between the two maxima amplitudes, entropy between these two states varies according to $t$. Taking the time derivative of result (27), one obtains

$$T\dot S = ml^2\dot\theta\ddot\theta + mgl\sin\theta\dot\theta, \tag{31}$$

becoming clear that the values of entropy vary according to $t$ due to the processes of energy transfer related to the conservative force, as represented in Figure 1(b). From $t = t_0$, entropy of the system increases ($t = 0$ s and $\dot{S}_0 = 0$) up to its maximum value ($\dot{S} = 15$ J rad$^2$/Ks), before the pendulum achieves its maximum velocity point at $\theta = 0$ rad. Notice $\theta = 0$ rad is not an equilibrium point ($E \neq 0$, reversible system) since the pendulum motion does not cease, and $\dot{S}$ continues decreasing up to reach its minimum value ($\dot{S} = -15$ J rad$^2$/Ks) at the symmetrical point of the initial position. Thus, entropy generated during the "going" phase of the pendulum is cancelled by the "returning" one, indicating a null entropy produced. At $t \approx 0.7$ s, the pendulum returns to its initial position, and the time rate of change in entropy reaches its maximum value again at $t \approx 0.9$ s, and minimum at $t \approx 1.1$ s, up to the moment when the pendulum returns to its initial position at $t \approx 1.4$ s. The values of $t$ where $\dot{S} = 0$, according to the equation (31), correspond to the points given by the solution of the differential equation (22), however, under different initial conditions, given that for $t = 0$ s we have $\theta_{max}$ and $\dot{S} = 0$.

### B. Non-Conservative Force

From the thermodynamic point of view, an irreversible system can be obtained by considering non-conservative forces, which result in changes in the energy balance of the system. Consequently, the quality of energy that can be used to produce work deteriorates due to dissipation, preventing the pendulum to return to its initial position. In this case, FLT is written as

$$T\Delta S_{irrev} > \Delta K + \Delta U_{grav} - W_{irrev},$$

and replacing the corresponding values, one obtains the inequality that furnishes a lower bound for the entropy when the damping force is given by equation (23)

$$T(S_B - S_A)_{irrev} > \frac{1}{2}ml^2(\dot{\theta}_B^2 - \dot{\theta}_A^2) + mgl(\cos\theta_A - \cos\theta_B) + bl^2\dot{\theta}(\theta_B - \theta_A). \quad (32)$$

Due to energy dissipation introduced by the non-conservative force (last term on the right-hand side of inequality (32)), $\dot{S}_{gen} > 0$ and, therefore, $T(S_B - S_A) > 0$. This energy dissipation term implies the turning points are never the same, resulting in the maximum "heights" that the pendulum can reach always decrease with each period. Therefore, the possible values for entropy during the "going" and "returning" motion are never the same, which does not allow for the cancellation of entropy produced in each period. The sum of non-canceled "excesses" grows, implying an entropy increase and a decrease in the energy available for producing work, which is manifested by a diminishing in the range of the pendulum motion. Thus, the values of maximum entropy are given by the points of return, which in turn will cease when the condition

$$\frac{1}{2}ml^2(\dot{\theta}_B^2 - \dot{\theta}_A^2) + mgl(\cos\theta_A - \cos\theta_B) > -bl^2\dot{\theta}(\theta_B - \theta_A), \tag{33}$$

cannot longer be satisfied. It occurs when the mechanical energy of the system is lower than the energy dissipated by the work related to the non-conservative force. Notice the right-hand side of the above inequality is positive since $\theta_A > \theta_B$, thanks to the damping force.

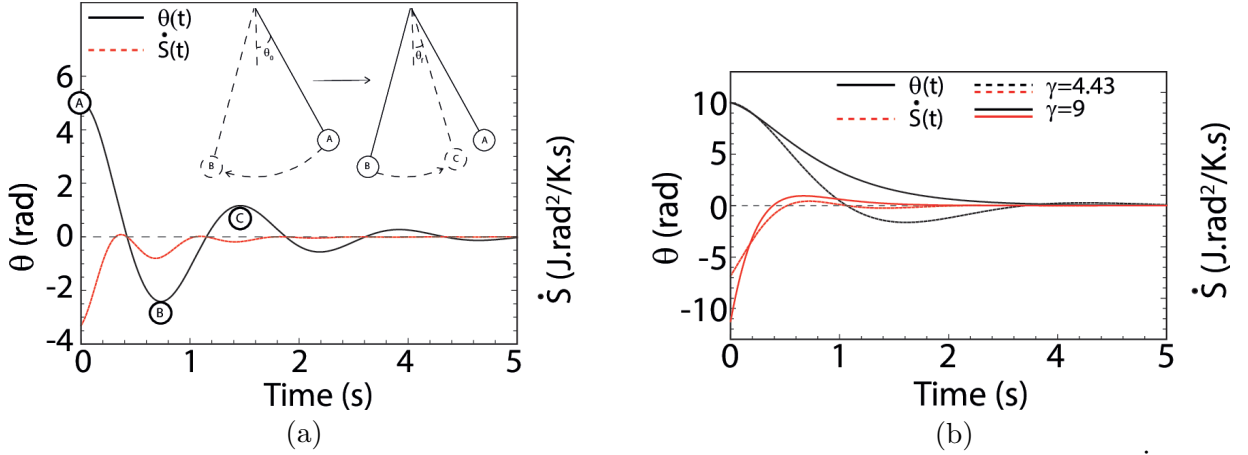

FIG. 3: Panel (a) shows the representation of the damped oscillation, and the curves for $\dot{S}(t)$ and $\theta(t)$. Points A, B, and C represent the initial, return and final position points, respectively. In panel (b), one has $\dot{S}(t)$ and $\theta(t)$ for the cases of critical damping ($\gamma = 4.43$) (dashed lines) and supercritical damping ($\gamma = 9$) (lines full). For calculation purposes, $g = 9.81$ m/s$^2$, $m = 1$ kg, $l = 0.5$ m, $T = 300$ K and $\omega = 4.43$ rad/s.

Considering two states with differential $d\theta$ e $d\dot{\theta}$, the increment $\Delta S_{AB} = S_B - S_A$ can be replaced by $dS$, resulting in

$$TdS > ml^2\dot{\theta}d\dot{\theta} + mgl\sin\theta d\theta + bl^2\dot{\theta}d\theta > 0. \tag{34}$$

Using the same procedure applied in (29), one obtains from the above inequality the following result

$$\dot{\theta}[\ddot{\theta} + \omega^2\theta + \gamma\dot{\theta}] + \gamma\ddot{\theta}\theta > 0, \tag{35}$$

where $\gamma = b/m > 0$. Observe that the term $\gamma\ddot{\theta}\theta > 0$ since $\theta$ is positive counterclockwise, implying the signs of $\dot{\theta}$ and $\ddot{\theta}$ are well-defined, possessing the same sign. The analysis of term $\dot{\theta}[\ddot{\theta}+\omega^2\theta+\gamma\dot{\theta}]$ is analogous to the one performed for the conservative case, given by the result (30). Notice, however, the term in the bracket corresponds to the damped pendulum. Then, one has

$$\ddot{\theta} + \omega^2\theta + \gamma\dot{\theta} = 0, \tag{36}$$

indicating the equation for the damped case can be given by the thermodynamic analyzis of the system.

Therefore, excluding the situations where $\dot{\theta} \neq 0$ rad/s as well as $\theta = 0$ rad, one rewrites the inequality (34) as

$$\left(\frac{\partial\dot{\theta}}{\partial\theta}\right) + \omega^2\left(\frac{\theta}{\dot{\theta}}\right) + \gamma > 0. \tag{37}$$

From a thermodynamic point of view, when one excludes symmetric points in a reversible system does not change the fact that entropy variation is null or constant in the system. For the pendulum without damping case, one can restrict our analysis to the situation where $\dot{\theta} = 0$ (symmetric points) are removed from the problem. Furthermore, at $\theta = 0$, we have only the vertical component of the weight force, implying that is an inversion point in the direction of the components of this force. Therefore, entropy is zero at this point, which means it can be excluded from the analysis of the equation (29). Hence, using the equation (29) and excluding the cases where $\dot{\theta} \neq 0$ and $\theta = 0$, one can write

$$\left(\frac{\partial\dot{\theta}}{\partial\theta}\right)_S = -\omega^2\left(\frac{\theta}{\dot{\theta}}\right), \tag{38}$$

where the subscript $S$ is used to indicate the isentropic process and, therefore, one can consider the reversible case as an adiabatic approximation. The result (38) may be read as follows: the rate of change of the angular velocity with position is equal to the ratio of position to velocity, and the constant of proportionality is the natural frequency of oscillation.

Using the above result, one can write the inequality (37) as

$$\left(\frac{\partial\dot{\theta}}{\partial\theta}\right)_{irrev} > \left(\frac{\partial\dot{\theta}}{\partial\theta}\right)_S - \gamma \tag{39}$$

Thus, the damping term $\gamma$, an irreversibility factor, leads to a smaller variation in the angular rate of velocity. Hence, pendulum motion has a smaller angular amplitude (lower energy) and velocity, as seen in Figure 1.The maximum (or minimum) points of the result (39) will no longer correspond to the maximum (and minimum) velocity points, being shifted due to the damping factor, as shown in Figure 4. For the cases of critically damping, the delay in the $\dot{\theta}$ curve tends to zero the higher the $\gamma$ and, consequently, the $-\omega^2(\dot{\theta}/\theta)$ and $\dot{\theta}$ approaches asymptotically each other until the oscillations ceases.

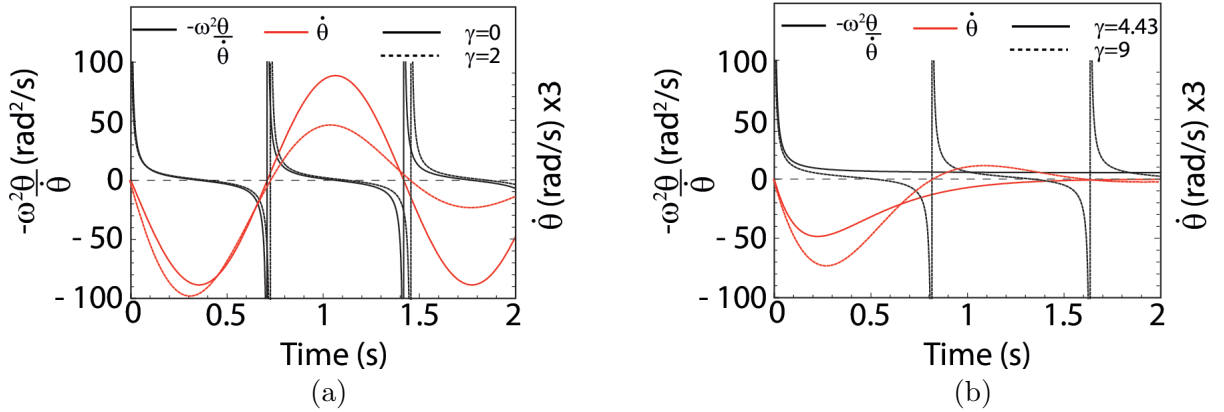

(a) (b)

FIG. 4: For the reversible case, panel (a) shows the maximum and minimum velocity, which corresponds to the points where $\left(\frac{\partial \dot{\theta}}{\partial \theta}\right) > 0$. For the irreversible case, panel (b), the damping factor shifts these points concerning the maximum and minimum points of the velocity. We use here $g = 9.81$ m/s$^2$, $m = 1$ kg, $l = 0.5$ m e $\omega = 4.43$ rad/s.

## VI. ENTROPY GENERATION FROM THE DAMPED PENDULUM

From the results obtained in the previous section, it is possible to quantify the entropy production rate for the mechanical system composed of the simple pendulum. From the relations (21), (27), and (34) the entropy generation $\dot{S}_{gen}$ is written, based on the Gouy-Stodola theorem, as

$$\dot{S}_{gen} = \frac{bl^2}{T}(\dot{\theta}^2 + (\theta - \theta_0)\ddot{\theta}), \qquad (40)$$

corresponding to entropy produced in the reversible system subtracted from the entropy produced in the irreversible one. Then, entropy produced corresponds to the time derivative of energy dissipation (power dissipated), being a result of the non-conservative force. Hence, energy dissipated is responsible by the entropy produced in the system.

Figure 3(a) shows the entropy variation as the $t$ flows (red dashed line). When the pendulum leaves the initial position, indicated by A, its entropy increases to a maximum value just before the pendulum reaches its maximum velocity at $\theta = 0$ rad, achieving a new minimum value before the pendulum reaches the return position, point B. At this time, entropy increases again until the pendulum reaches the new position C. The same analysis can be done considering the critical damping ($\gamma = 4.48$ and dashed lines in Figure 3(b)), and critically ($\gamma = 9$ and solid lines in Figure 3(b)).

In analogy to a heat engine, this part of the energy corresponds to the heat flow that the system must release to the cold reservoir to satisfy the entropy conditions established by the SLT. Such conditions obey the condition indicated by the equation (40) and expressed by

$$\dot{S}_{gen} > 0 \rightarrow (\dot{\theta}^2 + (\theta - \theta_0)\ddot{\theta}) > 0,$$

and, therefore,

$$\dot{\theta}^2 > -(\theta - \theta_0)\ddot{\theta}, \tag{41}$$

which is satisfied for all $\theta \leq \theta_0$ since the existence of the non-conservative force implies a loss of energy and a consequent decrease in the amplitude of oscillation.

For illustrative purposes only, Figure 5 represents the entropy generation $\dot{S}_{gen}$ for the damped systems considering different values of $\gamma$. For all damping cases, $\dot{S}_{gen}$ increases from $t = 0$ s, reaching a point of maximum value which approximately corresponds to the maximum velocity. Considering the inversion of the values of angular position, velocity, and acceleration, $\dot{S}_{gen}$ initiates an oscillatory motion whose amplitude decreases, vanishing when the pendulum ceases its motion. From the result (40), the dependence of $\dot{S}_{gen}$ on the factor $\gamma \, (= b/m)$ becomes evident. Thus, the greater the damping factor, the lower the amplitude of oscillation and, hence, the lower the energy available in the system. Therefore, the greater the dissipated energy, the greater the entropy generated in the system.

## VII. FINAL REMARKS

The First and Second Laws of Thermodynamics play a fundamental role in physics. These laws concern both the availability of energy and whatsoever form this energy can be used to produce useful work in that system. Although FLT is easy to understand, its application seems to be restricted only to thermal systems in graduate and secondary education textbooks.

Contrary to FLT, related to a conservation principle, the SLT, widely disseminated from the definition of Kelvin and Planck, establishes the increase of entropy for every real physical system, i.e. taking into account the presence of irreversibility. From the Gouy-Stodola theorem, the relation between both laws becomes clear, which occurs through the computation of the work due to conservative (reversible process) and non-conservative forces (irreversible process).

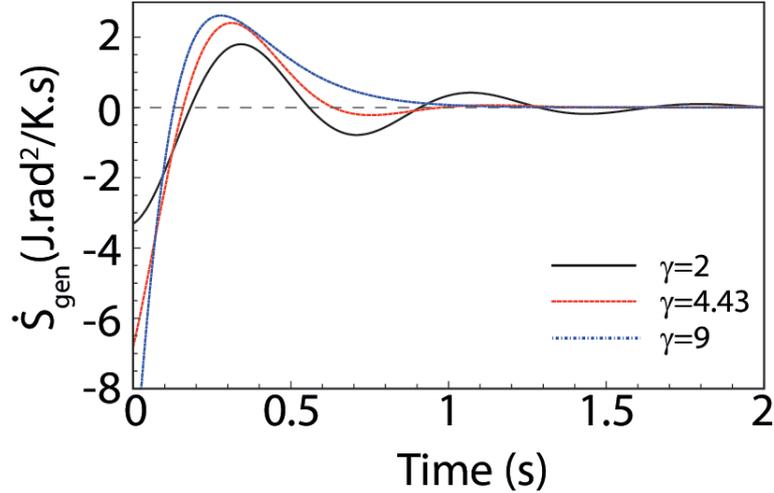

FIG. 5: Entropy production rate, $\dot{S}_{gen}$, considering the different damping cases. We use $b = \gamma m$, $m = 1$ kg, $l = 0.5$ m e $T = 300$ K.

For ideal systems, although the entropy variation between any two states is zero, the entropy production varies between the possible states along the path. It is important to keep in mind the difference between the definitions of entropy and entropy production: Entropy can be calculated for any physical system, and it is an extensive property of the system (state function) and, therefore, it is process-independent. As the dissipative force characterizes entropy production, entropy production is process-dependent. Both entropy and entropy production exists for reversible and irreversible processes. However, in the first process both remain constant while not for the latter. In the simple pendulum case, the irreversibility is introduced by a non-conservative force, which causes a decrease in the energy content due to dissipation. By the Gouy-Stodola theorem, the dissipated energy represented by the work concerning the non-conservative force is closely related to the entropy generation of the system. From the thermodynamic point of view, the lower amplitude is related to the lower entropy values. This result prevents the system to reach its initial energy. Then, at each period, the angular variation decreases, tending to zero in a finite time interval.

Finally, notice that energy dissipation turns the simple pendulum into a good mechanical example of applying the laws of thermodynamics. Ideally, the laws of mechanics should correspond to the laws of thermodynamics. However, correctly characterizing a mechanical system from the perspective of thermodynamics is not a simple task, requiring a detailed analysis of the system under consideration.

## Acknowledgment

RHL and SDC thanks to UFSCar.

---